\documentclass{PoS}
\usepackage[dvips]{epsfig}
\def\lsim{\raise0.3ex\hbox{$<$\kern-0.75em\raise-1.1ex\hbox{$\sim$}}}
\def\gsim{\raise0.3ex\hbox{$>$\kern-0.75em\raise-1.1ex\hbox{$\sim$}}}

\title{Recent lattice results on finite temperature and density QCD, part II}

\ShortTitle{QCD thermodynamics II}

\author{\speaker{Frithjof Karsch}\thanks{This work has been supported in
part by contract DE-AC02-98CH10886 with the U.S. Department of Energy.}\\
  Physics Department, Brookhaven National Laboratory, Upton, NY 11973, USA\\
      E-mail: \email{karsch@bnl.gov}}

\abstract{We discuss recent progress in studies of QCD thermodynamics
with almost physical light quark masses and a physical value of the 
strange quark mass. We summarize results on the transition
temperature in QCD and analyze the relation between deconfinement and
chiral symmetry restoration.
}

\FullConference{The XXV International Symposium on Lattice Field Theory\\
		 July 30-4 August 2007\\
		 Regensburg, Germany}

\begin{document}

\section{Introduction}

The QCD vacuum has a complicated structure that manifests itself 
most prominently in the confinement of quarks and gluons and the  
spontaneous breaking of chiral symmetry. These non-perturbative
properties of QCD are known to be temperature dependent and eventually
will disappear at high temperature; at least in the limit of vanishing
quark masses we expect that QCD undergoes a phase transition from
a hadronic phase to a new phase of deconfined quarks and gluons in
which chiral symmetry is restored. It has been speculated
that there could be two separate transitions in QCD at which quarks
and gluons deconfine and chiral symmetry gets restored 
\cite{Shuryak,Hatta1}, a scenario that has indeed been observed in lattice
calculations of gauge theories with fermions in the adjoined representation
\cite{aQCD}. In QCD with quarks being in the fundamental representation,
however, it seems that at least for vanishing quark chemical potential
there is a unique transition in the chiral limit at which quarks and gluons 
deconfine and chiral symmetry gets restored\footnote{Large $N_c$ arguments
suggest that the situation could be more complicated for non-zero quark
chemical potential \cite{McLerran}.}.

While in massless QCD the chiral condensate is a unique order parameter for 
chiral symmetry restoration, there is no counterpart for deconfinement.
The Polyakov loop, which is an order
parameter for deconfinement in the limit of infinitely heavy quarks
\cite{Kuti}, is non-zero at all values of the temperature whenever quarks
have finite masses. Nonetheless, the deconfining properties of the QCD
transition are clearly reflected in the behavior of bulk thermodynamic
observables, e.g. in the rapid rise of the energy or entropy density as 
well as in the sudden increase in fluctuations of light and strange 
quark numbers. The  sudden change in the latter 
reflects the liberation of many light degrees of freedom,
quarks and gluons, which dominate the properties of the thermal medium at
high temperature. In the chiral limit these sudden changes  
go along with singularities in bulk thermodynamic observables, the 
specific heat as well as quartic fluctuations of the light quark number
diverge or develop a cusp. At the same temperature the chiral order 
parameter and its derivative with respect to the quark mass, the chiral 
susceptibility, show singular behavior. In this limit it is obvious that 
the singular behavior in observables related to deconfinement and 
chiral symmetry restoration, respectively, are closely related. To what
extent this close relation persists also for non-zero values of the
quark masses then becomes a quantitative question that should be answered
through numerical calculations in lattice QCD. 

This became of particular interest
in view of a recent calculation \cite{aoki_Tc} that suggested that 
there might be a large difference in the transition temperature related
to deconfinement observables on the one hand and observables sensitive
to chiral symmetry restoration on the other hand. It has been suggested
that in the continuum limit this difference can be as large as $25$~MeV.  
However, calculations performed with ${\cal O}(a^2)$ improved staggered 
fermions \cite{p4_Tc,lat07} so far did not show such a large difference. 

In this write-up we will discuss some results from lattice calculations
concerning the interplay of deconfinement and chiral symmetry restoration. 
As these
results have been presented in July/August of 2007 in a very similar
format at the '4th International Workshop on Critical Point and Onset of
Deconfinement' and at the 'XXV International Symposium on Lattice Field
Theory' the write-up of these talks has been splitted into two parts.
In the first part \cite{partI} we discussed recent results on the QCD
equation of state.  In this second part we will concentrate on results that
can give insight into properties of the QCD transition itself. 
We will focus here on a presentation of results obtained with 
${\cal O}(a^2)$ improved staggered fermion formulations. Results obtained
with the 1-link, stout smeared staggered fermion action \cite{aoki_Tc} have 
been presented at both meetings separately \cite{Fodor}.

\section{The transition temperature}

Before entering the discussion on deconfinement and chiral symmetry restoration,
let us briefly summarize
the current status of calculations of the QCD transition temperature
using various discretization schemes. The goal here is, of course, to
determine the transition temperature in the continuum limit of lattice
regularized QCD with its
physical spectrum of two light and a heavier strange quark mass.
While the heavier quarks, e.g. the charm quarks, may
influence thermodynamics at high temperature \cite{laine,michael}, they are 
not expected to affect the transition temperature. In fact, even dynamical 
strange quark degrees of freedom seem to have little influence on the value of 
the transition temperature. Differences in the transition
temperatures of $2$, $(2+1)$-flavor and $3$-flavor QCD still seem to be well 
within the current statistical and systematic uncertainty.
This is in agreement with the observed weak dependence of the transition 
temperature on the light quark mass or, equivalently, on the light 
pseudo-scalar meson mass,
\begin{equation}
r_0 T_c(m_{PS}) -r_0 T_c(0) \simeq A \left( r_0 m_{PS}\right)^{d} \; ,
\label{Tc_m_scaling}
\end{equation}
with $d\simeq 1$, $A\lsim 0.05$. To be specific we have used here the distance
$r_0$ extracted from the static quark potential (see part I \cite{partI}) to 
set the scale for $T_c$. 
The weak quark mass dependence of $T_c$ is consistently 
found in calculations with ${\cal O}(a^2)$ improved staggered 
fermions \cite{p4_Tc,milc_Tc,old_p4_Tc} as well as with Wilson 
fermions \cite{Bornyakov1,Bornyakov2,Maezawa}. 
This has been taken as an indication for the importance of a large number
of rather heavy resonances for building up the critical conditions,
e.g. a sufficiently large energy density, needed to deconfine the partonic
degrees of freedom in QCD. As these heavy resonances are
only weakly dependent on the quark mass values the light chiral sector
of QCD may play a subdominant role for the quantitative value of the
transition temperature, while it does, of course, control the universal
properties of thermodynamic observables in the chiral limit.
In 2-flavor QCD for instance\footnote{The discussion carries over to
the light quark sector of (2+1)-flavor QCD. However, in this case one
has to keep in mind that a second order transition point
might be reached already at a non-zero value of the light quark mass.
This effectively will change the universality class of the transition
from $O(4)$ to $Z(2)$ and modifies the singular structure of the free 
energy.}, the scaling exponent $d$ 
appearing in Eq.~\ref{Tc_m_scaling}, will be related to critical 
exponents ($\beta$, $\delta$) of $3$-dimensional, $O(4)$ symmetric spin models,
$d=2/\beta\delta = 1.08$.    

\begin{figure}[t]
\epsfig{file=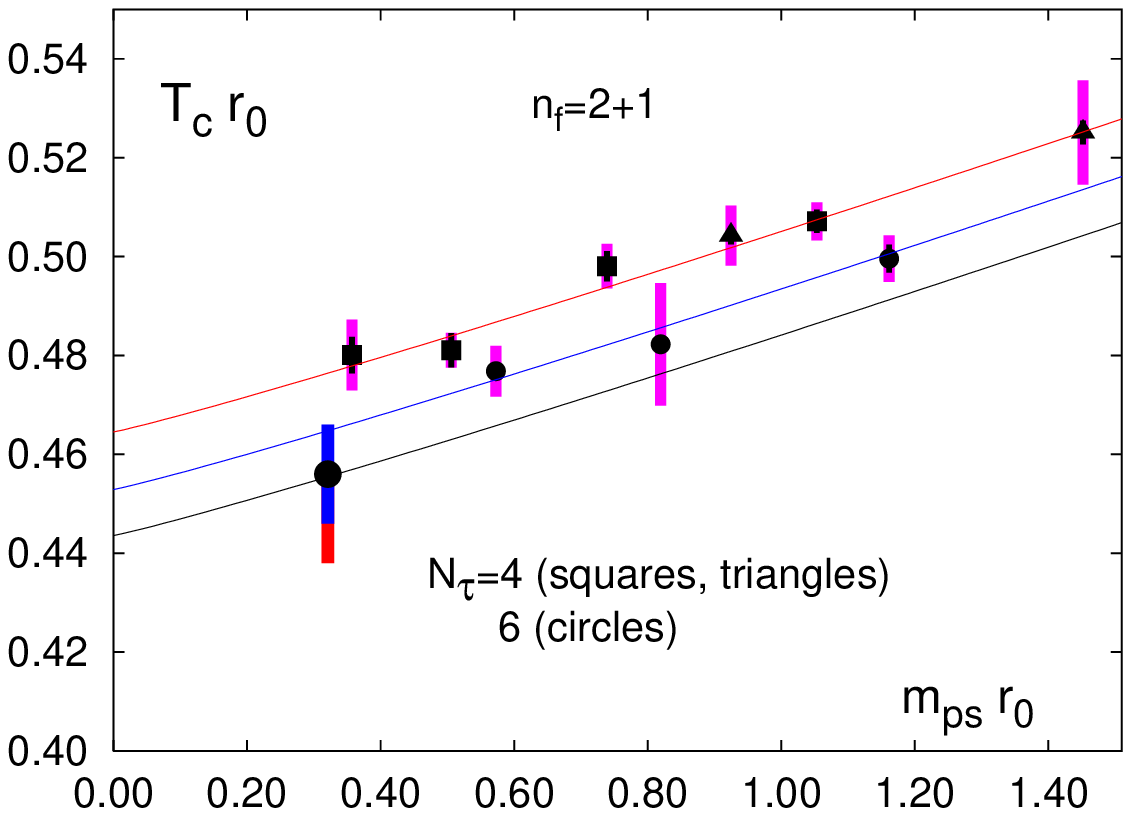,width=8.0cm}
\epsfig{file=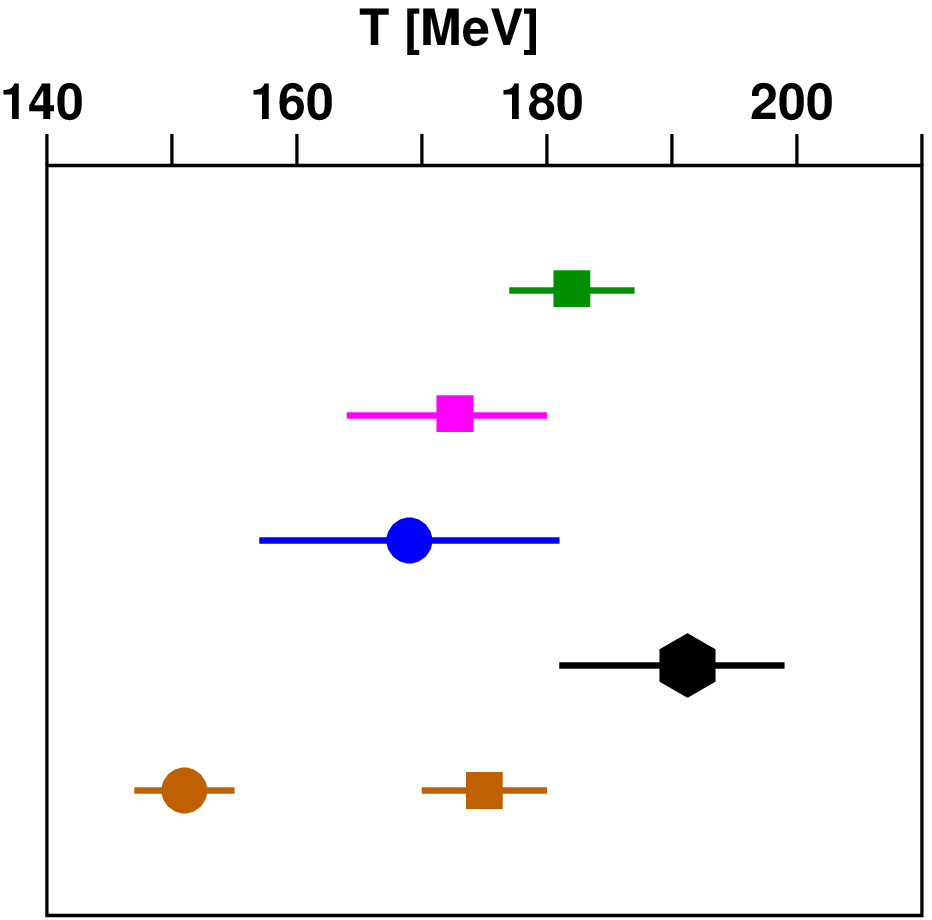,width=7.0cm}
\caption{Left: Quark mass and cut-off dependence of the transition temperature 
calculated with the p4fat3 staggered fermion action on lattices with temporal 
extent $N_\tau=4$ and $6$ \cite{p4_Tc}.
Right:
Transition temperatures determined in several recent studies
of QCD thermodynamics. From top to bottom the first two data points show 
results obtained in simulations of 2-flavor QCD using clover improved 
Wilson fermions on lattices with temporal extent $N_\tau =8,\; 10$ and $12$ 
\cite{Bornyakov1,Bornyakov2} and $N_\tau =4$ and $6$ 
\cite{Maezawa}, respectively. The remaining data points have been obtained
in simulations of QCD with 2 light quark masses and a physical strange
quark mass. They are based on calculations with staggered fermions using the 
asqtad action on $N_\tau =4$, $6$ and $8$ lattices \cite{milc_Tc},
the p4fat3 action on $N_\tau =4$, $6$ \cite{p4_Tc} and 1-link, stout 
smeared action on $N_\tau =4$, $6$, $8$ and $10$ lattices \cite{aoki_Tc}.
Circles indicate that the determination of the transition
temperature is based on observables sensitive to chiral symmetry restoration,
i.e. the chiral condensate and susceptibilities deduced from it. Squares
indicate that observables sensitive to deconfinement have been used to
determine the transition temperature, e.g.the Polyakov loop, its susceptibility
and/or light and strange quark number susceptibilities. The diamond indicates
that both sets of observables have been analyzed. With the exception of results
presented in \cite{Maezawa} all calculations aimed at an extrapolation
to the continuum limit ($N_\tau \rightarrow \infty$) for physical values
of the quark masses. All results have
been rescaled to a common physical scale using $r_0=0.469$~fm \cite{Gray}.}
\label{fig:Tc_summary}
\end{figure}

In Fig.~\ref{fig:Tc_summary}(left) we show results on the quark mass 
dependence of the
transition temperature obtained in calculations with the p4fat3 staggered
fermion action on lattices with two different values of the cut-off,
$aT=1/4$ and $1/6$ \cite{p4_Tc}.
As expected, in addition to the obvious quark
mass dependence of $T_c$ also a dependence on the cut-off, $a$, is clearly 
visible. Asymptotically the cut-off dependence is expected to be proportional 
to $a^2$, {\it i.e.} we expect to find  
\begin{equation}
r_0 T_c(m_{PS}, N_\tau) -r_0 T_c(0,\infty) 
\simeq A \left( r_0 m_{PS}\right)^{d} +B/N_\tau^2\; .
\label{Tc_m_cont_scaling}
\end{equation}
This ansatz generally is used to extrapolate to the continuum limit and 
to extract the transition temperature, $T_c \equiv T_c(0,\infty)$. Of course,
when using Eq.~\ref{Tc_m_cont_scaling} for an extrapolation to the continuum 
limit one has to make sure that the asymptotic scaling regime has been
reached.
In Ref.~\cite{p4_Tc} the extrapolation is only based on two different
values of the lattice cut-off, $aT=1/4$ and $1/6$, which may not be
close enough to the continuum limit. This has been taken into account 
in the analysis performed in Ref.~\cite{p4_Tc} by estimating a systematic 
error for the possible scaling violations. This lead to an estimate of the 
transition temperature $T_c = 192(7)(4)$~MeV with the second error denoting
an estimate for the systematic uncertainty in the extrapolation. 
An earlier analysis performed with the asqtad action
on lattices with temporal extent $N_\tau=4$, $6$ and $8$ but 
smaller spatial volume, $N_\sigma/N_\tau =2$, 
lead to the estimate
$T_c = 169(12)(4)$~MeV \cite{milc_Tc}. Both calculations currently get 
improved in a systematic comparison of simulations performed with the p4fat3 
and asqtad action on lattices of size $32^3\, 8$ \cite{lat07}. 

\begin{figure}[t]
\begin{picture}(162,144)
\put(5,7){\epsfig{file=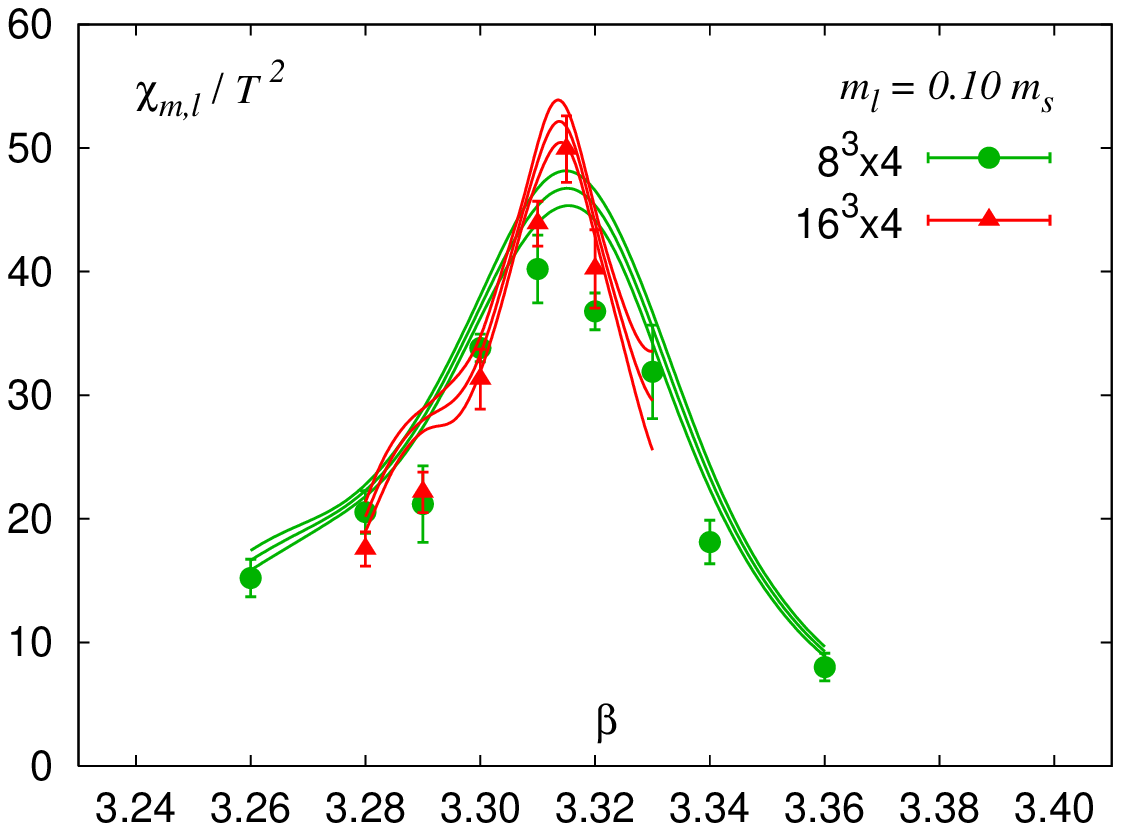,width=7.5cm}}
\put(5,11){\epsfig{file=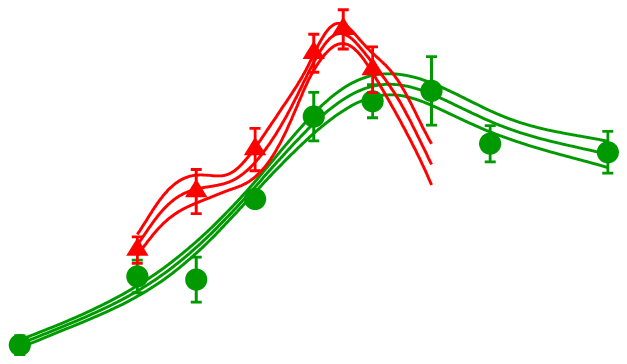,width=7.35cm,
height=3.5cm}}
\put(205,7){\epsfig{file=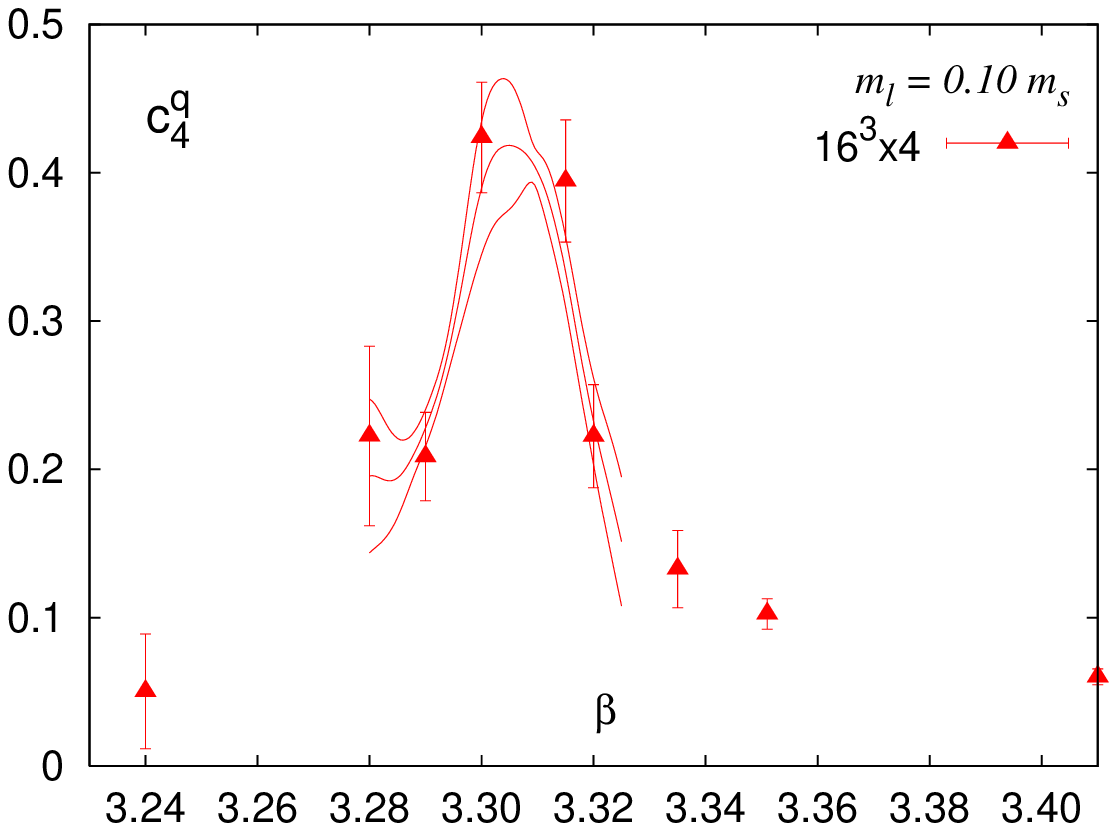,width=7.5cm}}
\end{picture}
\caption{Disconnected part of the light quark chiral susceptibility 
and the Polyakov loop susceptibility (left) \cite{p4_Tc} and the quartic 
fluctuations of the light quark number (right) \cite{Schmidt} calculated 
on lattices with temporal
extent $N_\tau =4$ in simulations with the p4fat3 action.}
\label{fig:sus46}
\end{figure}

In calculations with the p4fat3 action Polyakov loop and
chiral susceptibilities have been examined. The transition temperature has
been determined by locating peaks in these susceptibilities 
(see Fig.\ref{fig:sus46}(left)). This lead
to consistent results on larger volumes, although systematic deviations
have been observed for smaller spatial volumes, e.g. for $N_\sigma/N_\tau =2$.
Calculations performed with the 1-link, stout smeared action \cite{aoki_Tc}
on lattices with temporal extent ranging from $N_\tau=4$ up to $N_\tau=10$, 
on the other hand, suggest that there exist differences in the location of 
the peak positions in chiral susceptibilities and inflection points 
determined for observables like the
Polyakov loop and strange quark number susceptibility.  These differences
seem to become 
increasingly significant with increasing $N_\tau$. In this analysis a 
transition temperature related to chiral properties is determined to be
$T_c = 151(3)(3)$~MeV while observables related to deconfinement suggest a 
transition temperature $T_c \simeq 175$~MeV. For further discussion of these 
calculations see also \cite{Fodor}. 

In the Wilson formulation a discussion of chiral symmetry restoration
becomes more involved than in the staggered case. For this reason 
only observables related to deconfinement, {\it i.e.} the Polyakov loop
and its susceptibility, have generally been analyzed in studies with Wilson
fermions. 
Compared to calculations of the QCD transition temperature performed with 
staggered fermions in ($2+1$)-flavor
\cite{aoki_Tc,p4_Tc,milc_Tc} and 3-flavor \cite{p4_3f} QCD, 
calculations with clover-improved Wilson fermions 
\cite{Bornyakov1,Bornyakov2,Maezawa} performed in 2-flavor QCD 
typically use rather large quark masses, $m_{PS}r_0\gsim 1$. This makes
an extrapolation to physical masses difficult. Nonetheless, a straightforward
application of the scaling ansatz, Eq.~\ref{Tc_m_cont_scaling}, for 
extrapolations of the transition temperature down to the physical values of 
the pion mass, $m_{PS}r_0 \simeq 0.32$, yields results that are in 
reasonably good 
agreement with values determined within staggered fermion formulations. 
The current status of calculations of transition temperatures with Wilson and 
staggered fermions is summarized in Fig.~\ref{fig:Tc_summary}(right).

\section{Deconfinement and chiral symmetry restoration}

The relation between deconfinement and chiral symmetry restoration in
QCD has been discussed since a long time \cite{Shuryak,Hatta1}. Although 
both phenomena
seem to be related to physics on different length scales lattice 
calculations seem to suggest that both phenomena happen at approximately
the same temperature even at finite, non-zero values of the quark masses
when none of the symmetries related to confinement ($Z(3)$ center symmetry) 
or chiral symmetry breaking ($SU_L(n_f)\times SU_R(n_f)$) are realized 
exactly. In calculations with almost physical light quark masses and 
a physical value of the strange quark mass this has recently been 
confirmed on lattices with temporal extent $N_\tau =4$ and $6$ in a 
detailed analysis performed with ${\cal O}(a^2)$ improved staggered fermions 
\cite{p4_Tc}. The gradual restoration of chiral symmetry with increasing
temperature is signaled by changes in the light and strange quark chiral
condensates and the corresponding susceptibilities,
\begin{equation}
\langle \bar{\psi}\psi \rangle_q = \frac{T}{V} 
\frac{\partial \ln Z}{\partial m_q} \;\; , \;\; \chi_{m,q} = \frac{T}{V} 
\frac{\partial^2 \ln Z}{\partial m_q^2} \; \; ,
\label{msus}
\end{equation}
where $q=l,\; s$ for the light and strange quark sector, respectively.
The onset of deconfinement, on the other hand, can be examined through
an analysis of the Polyakov loop $L$ and its susceptibility,
$\chi_L$,
\begin{equation}
L = \langle \frac{1}{3 N_\sigma^3} \sum_{\vec{n}}
{\rm Tr} \prod_{n_0=1}^{N_\tau} U_{(n_0,\vec{n}),\hat{0}} \rangle \;\; , \;\; 
\chi_L = N_\sigma^3 \left( \langle L^2\rangle -\langle L\rangle^2\right)\; ,
\label{deco}
\end{equation}
where $U_{(n_0,\vec{n}),\hat{0}}$ denote the gauge field variables defined
on temporal links of a lattice of size $N_\sigma^3 N_\tau$.
Some results for the light quark chiral susceptibility, $\chi_{m,l}$, and the 
Polyakov
loop susceptibility, $\chi_L$, calculated in a simulation with a physical
value of the strange quark mass and a light quark mass that corresponds
to a light pseudo-scalar mass of about $220$~MeV ($m_l/m_s=0.1$)
on lattices of size $16^3 4$ are shown in Fig.~\ref{fig:sus46}(left).

As can be deduced from this figure
the Polyakov loop susceptibility does not provide
a strong signal for deconfinement in calculations with light
dynamical quarks; the Polyakov loop itself is non-zero at
all temperatures and rises smoothly through the transition region. This
results only in a shallow peak in $\chi_L$, which nonetheless is in 
good agreement with the peak position in the light quark chiral susceptibility,
$\chi_{m,l}$.

In part I \cite{partI} we have presented results on bulk thermodynamic
observables, e.g. the energy density ($\epsilon/T^4$), as well as light 
and strange quark number susceptibilities ($\chi_{l,s}/T^2$). 
At least on lattices with temporal
extent $N_\tau =4$ and $6$ they both rise rapidly in about the 
same temperature range. In both cases this is due to the deconfining
nature of the QCD transition; as indicated in the introduction both
quantities are sensitive to the liberation of many light quark and
gluon degrees of freedom. In the chiral limit the sudden rise of,
e.g. $\epsilon/T^4$ and $\chi_{l,s}/T^2$ seems to be closely related to the 
singular behavior of the QCD partition function that arises from the 
restoration of chiral symmetry. To be specific let us discuss here the
chiral limit of 2-flavor QCD (see footnote 2).
The singular part of the free energy, $f_s$, is controlled 
by a reduced 'temperature' $t$ that is a function of temperature
as well as the quark chemical potential $\mu_q$ \cite{Hatta}. The latter adds  
quadratically to the reduced temperature in order to respect charge symmetry 
at $\mu_q=0$,
\begin{equation}
f_s(T,\mu_q) = b^{-1} f_s(tb^{1/(2-\alpha)})
\sim t^{2-\alpha} \;\; , \; {\rm with} \;\;
t= \left| \frac{T-T_c}{T_c}\right| +c 
\left(\frac{\mu_q}{T_c}\right)^2 \;\; ,
\label{free}
\end{equation}
where $b$ is an arbitrary scale parameter and $\alpha$ denotes a 
critical exponent.
As a consequence the specific heat as well as the quartic fluctuations
of the light quark number, 
$c_4^q \sim \left( \langle N_q^4\rangle - 3 \langle N_q^2 \rangle\right)$, 
show singular behavior at the critical point $t=0$ that is controlled by the 
same critical exponent $\alpha$,
\begin{equation}
C_V\sim \frac{\partial^2 \ln  Z}{\partial T^2} \sim t^{
{-\alpha}} \;\; ,\;\; 
c_4^q  \sim \frac{\partial^4 \ln  Z}{\partial \mu_{q}^4} \sim t^{
{-\alpha}} \;\; ,\; {\rm for ~~}\mu_q =0.
\label{critical}
\end{equation}
Unlike the Polyakov loop susceptibility the quartic fluctuations of the
quark number thus provide a strong signal for deconfinement. This is shown
in Fig.~\ref{fig:sus46}(right). In fact, a comparison of $c_4^q$ calculated
here in ($2+1)$-flavor QCD with light quark masses that correspond to a light 
pseudo-scalar (pion) mass of about $220$~MeV \cite{Schmidt} and earlier 
calculations in 2-flavor QCD with $10$ times heavier quarks 
corresponding to a pion mass of about $770$~MeV \cite{c6} show that 
the quartic fluctuations rise strongly with decreasing quark mass.

\begin{figure}[t]
\begin{center}
\epsfig{file=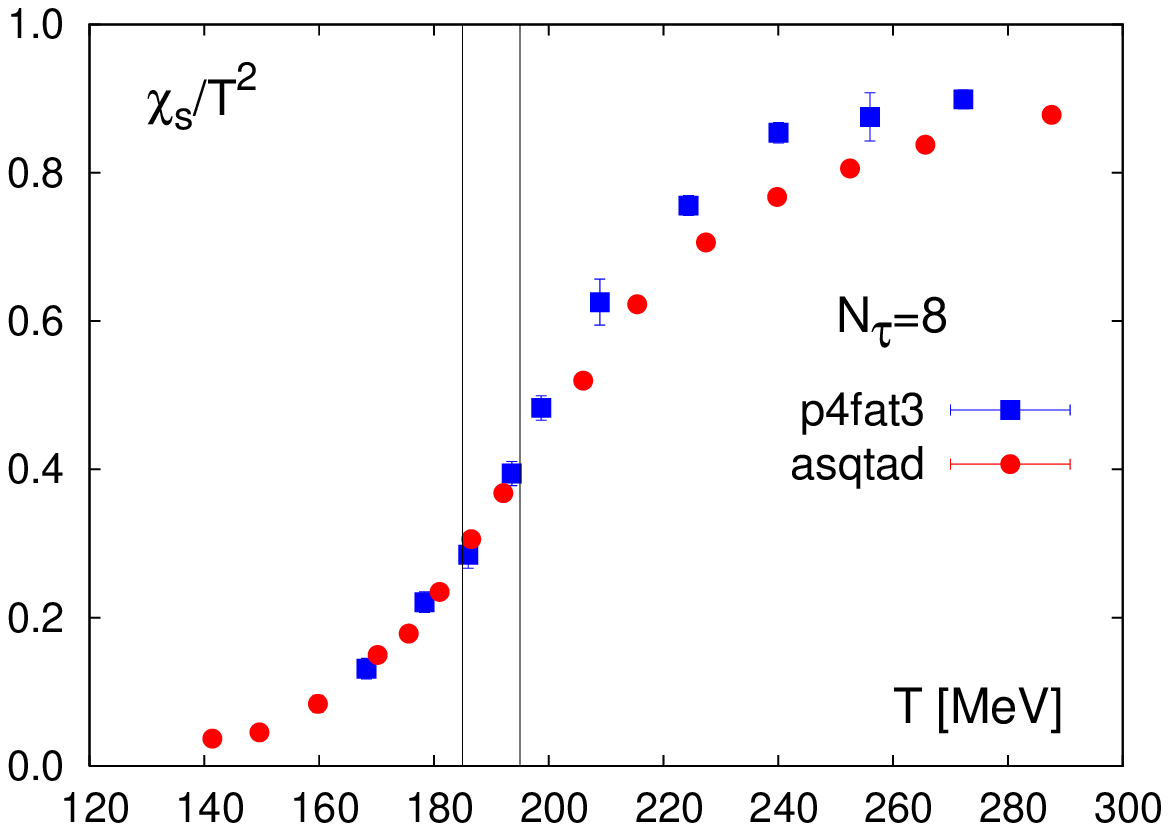,width=7.5cm}
\epsfig{file=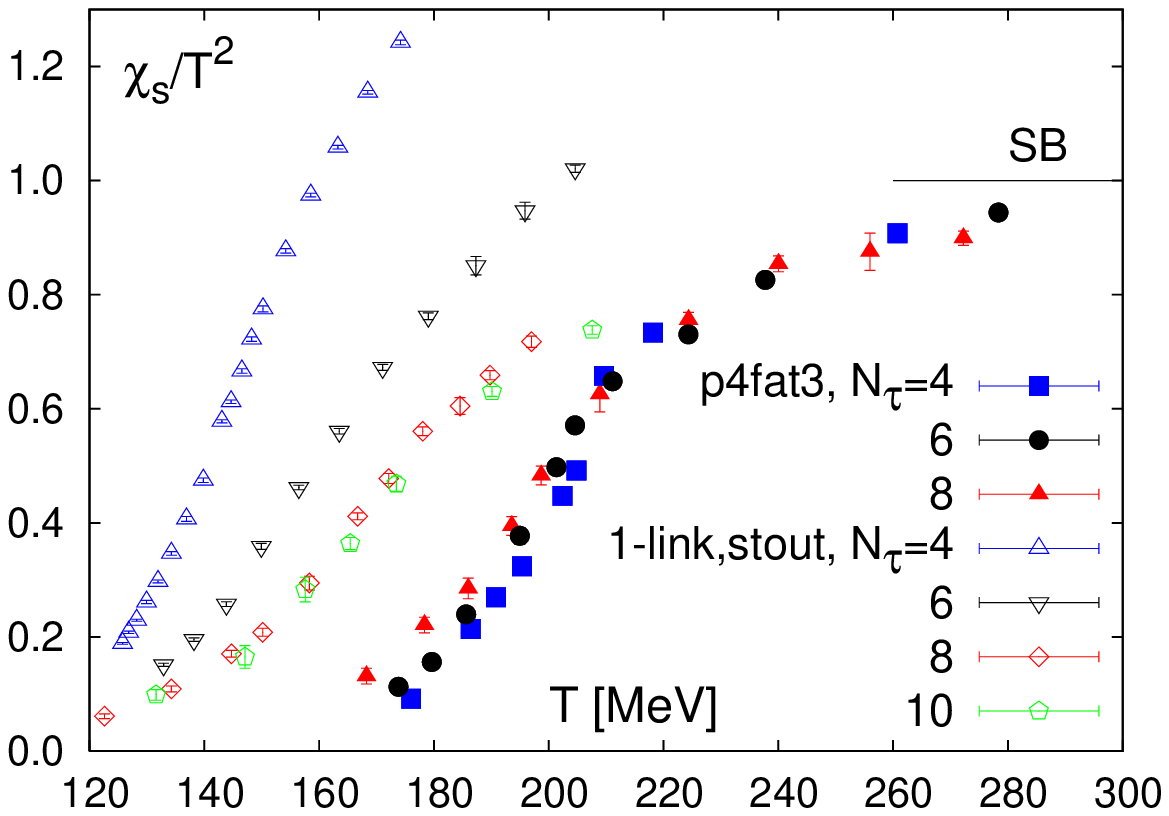,width=7.5cm}
\end{center}
\caption{Preliminary results of the hotQCD collaboration \cite{lat07}
for the strange quark number susceptibility calculated on lattices of
size $32^3 8$ using two different ${\cal O}(a^2)$ improved staggered fermion 
actions, asqtad and p4fat3 (left). The vertical lines indicate a band of
temperatures, $185{\rm MeV} \le T \le 195{\rm MeV} $, which characterizes the
transition region in the $N_\tau =8$ calculations \protect\cite{lat07}
(see also Figs. $4$ and $5$). The right
hand figure shows a comparison of calculations performed with the p4fat3
action on lattices of temporal extent $N_\tau =4$, $6$ \cite{Schmidt} and $8$ 
\cite{lat07} and with
the 1-link, stout smeared action for $N_\tau =4$, $6$, $8$ and $10$ 
\cite{aoki_Tc}.
Note that different conventions have been used to define the temperature scale
(see text)
}
\label{fig:sus8}
\end{figure}

In Fig.~5 of part I \cite{partI} we have shown results for the strange
quark number susceptibility, {\it i.e.} the fluctuations of strangeness
number $\chi_s \sim \langle N_s^2 \rangle$, calculated with the p4fat3
and asqtad actions on lattices of temporal extent $N_\tau =4$, $6$ 
\cite{milc_Tc,Schmidt} and $8$ \cite{lat07} in (2+1)-flavor QCD and 
a light to strange quark mass ratio $m_l/m_s=0.1$.
The $N_\tau =8$ results are preliminary results obtained by the hotQCD
collaboration. They are discussed in more detail in \cite{lat07}. 
These calculations indicate a quite good agreement
between results obtained with the two different ${\cal O}(a^2)$ improved 
discretization schemes, although in particular at temperatures above the 
crossover region some differences show up. This is more clearly seen in 
Fig.~\ref{fig:sus8}(left) where we compare the preliminary
results obtained within both discretization schemes on $N_\tau =8$ lattices. 
These differences may 
be due to small differences in the choice of quark masses that define
the constant line of physics along which the calculations have been 
performed and may partly be also due to
differences in the discretization errors for both actions
which may be about\footnote{In the infinite temperature
limit deviations from the continuum value, $\chi_{free, m\equiv 0}^{SB}$,
can be calculated analytically. For massless free staggered fermions on
lattices with temporal extent $N_\tau=8$ this yields
$\chi_{free, m\equiv 0}/ \chi_{free, m\equiv 0}^{SB} = 
0.92$ (asqtad), $0.98$ (p4fat3), $1.47$ (1-link,stout). 
Like in the case of the pressure and other bulk
thermodynamic observables, cut-off effects in the quark number susceptibility
are ${\cal O}(a^2)$ improved for the asqtad and p4fat3 action and only start 
with $1/N_\tau^4$ corrections in the infinite temperature limit.}
$6\%$ for $N_\tau=8$. These differences as well as the cut-off dependence
of results obtained on $N_\tau=4,~6$ and $8$ lattice with the asqtad
and p4fat3 actions are, however, small 
when compared to results obtained with the 1-link, stout smeared staggered 
fermion action \cite{aoki_Tc} as shown in Fig.~\ref{fig:sus8}(right).
The differences between the asqtad and p4fat3 calculations on the one hand
and the 1-link, stout smeared calculations on the other hand arise
from two sources. For small values of $N_\tau$,  the quark number 
susceptibilities calculated with 1-link staggered fermion actions overshoot
the continuum Stefan-Boltzmann result at high temperatures and reflect 
the strong cut-off dependence of thermodynamic observables calculated with
this action. This is well-known to happen in the infinite temperature, ideal 
gas limit and influences the behavior of thermodynamic observables in the
high temperature phase of QCD (see footnote 3 and also Fig.~2 in 
\cite{partI}). On the other hand the differences also arise from the 
different choice for the zero temperature observable used to set the 
temperature scale. While the temperature scale in the asqtad and p4fat3
calculations has been obtained from the static quark potential (the distance
$r_0$), the kaon decay constant has been used in calculations with the 
1-link, stout smeared action. Of course, this should not make a difference
after proper continuum extrapolations have been carried out. At finite
values of the cut-off, however, one should make an effort to disentangle
cut-off effects in thermodynamic observables from cut-off effects that
only arise from a strong lattice spacing dependence in a zero temperature 
observable
that is used to define a temperature scale. In this respect, the scale
parameter $r_0$ extracted from the heavy quark potential is a safe quantity
which is easy to determine; it has been studied in detail and its weak
cut-off dependence is well controlled \cite{p4_eos,asqtad_pot}. 

\begin{figure}[t]
\begin{center}
\epsfig{file=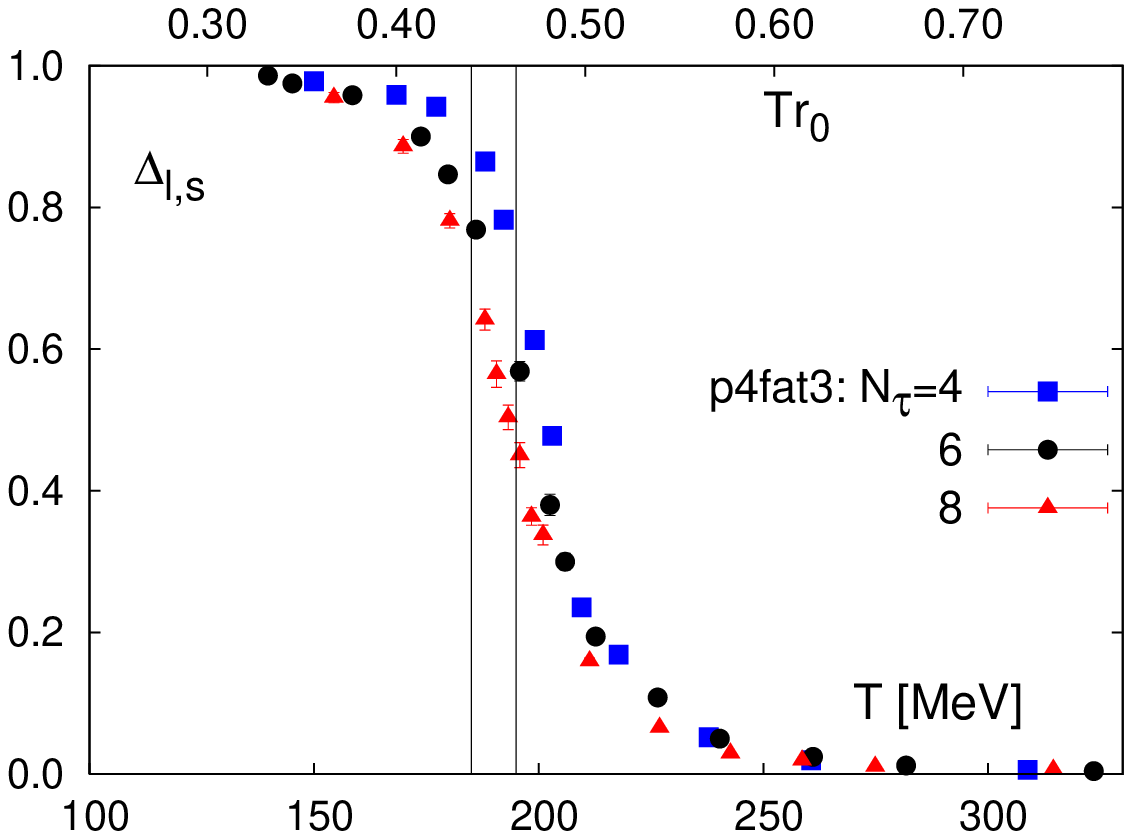,width=7.5cm}
\epsfig{file=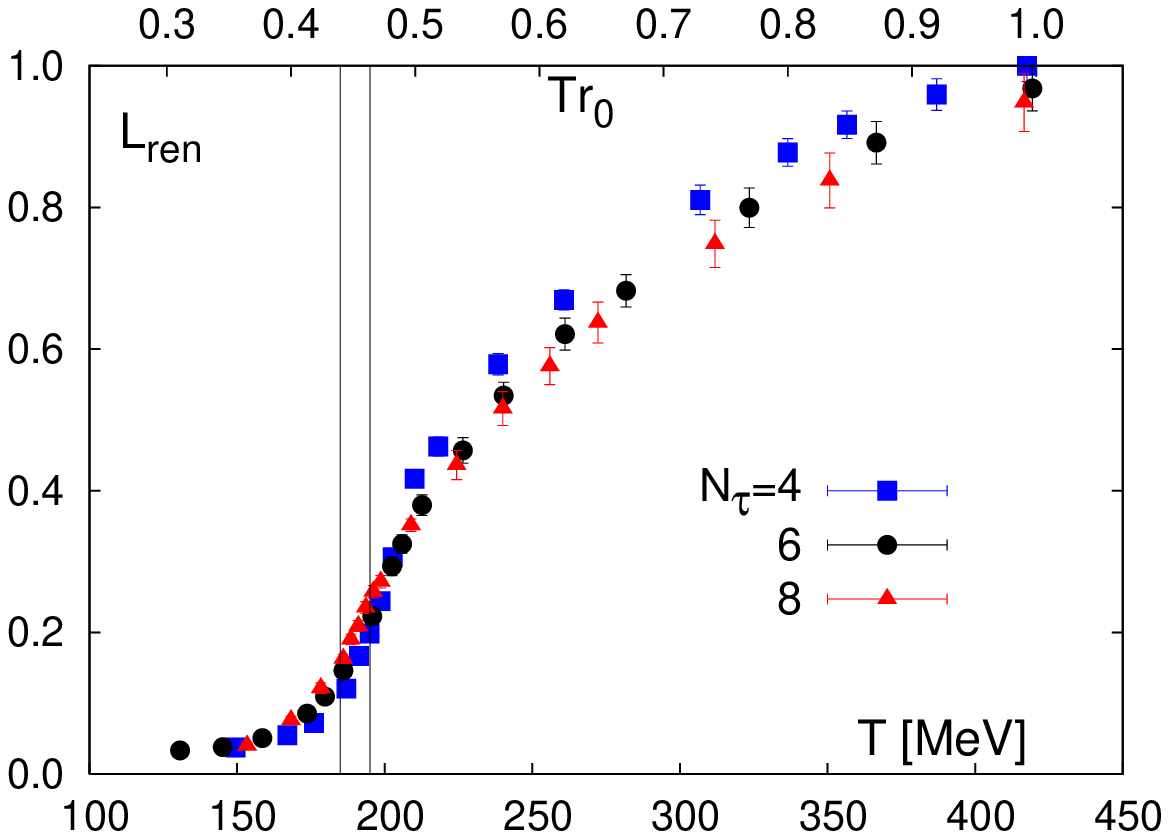,width=7.5cm}
\end{center}
\caption{The difference of light and strange quark chiral 
condensates normalized to its zero temperature value as defined in
Eq.~\protect\ref{delta} (left) and the 
renormalized Polyakov loop expectation value (right).
Shown are results from simulations on $N_\tau=4$ and $6$ lattice 
obtained with the p4fat3 \cite{p4_eos} action as well as 
preliminary results for $N_\tau =8$ obtained by the
hotQCD Collaboration \cite{lat07}. The upper axis shows the temperature
in units of the distance $r_0$ extracted from the heavy quark potential. The
lower temperature scale in units of MeV has been obtained from this using
$r_0 = 0.469$~fm \cite{Gray}. The vertical lines indicate a band of 
temperatures, $185{\rm MeV} \le T \le 195{\rm MeV} $, which characterizes the
transition region in the $N_\tau =8$ calculations.}
\label{fig:chipol}
\end{figure}

Let us now turn our attention to observables sensitive to chiral symmetry
restoration which, of course, is
signaled by changes in the chiral condensate (Eq.~\ref{msus}). This also
is reflected in pronounced peaks in the light quark chiral susceptibility as 
shown in Fig.~\ref{fig:sus46}. 
As the chiral condensate receives additive as well as multiplicative
renormalization, one should look at appropriate combinations that 
eliminate the renormalization effects. An appropriate choice is to subtract
a fraction of the strange quark condensate from the light quark condensate
and normalize the finite temperature difference with the corresponding
zero temperature difference,
\begin{equation}
\Delta_{l,s}(T) = \frac{\langle \bar{\psi}\psi \rangle_{l,T} -
\frac{m_l}{m_s}
\langle \bar{\psi}\psi \rangle_{s,T}}{\langle \bar{\psi}\psi \rangle_{l,0} -
\frac{m_l}{m_s} \langle \bar{\psi}\psi \rangle_{s,0}} \; .
\label{delta}
\end{equation}
We note, that the strange quark contribution to this quantity will drop
out in the chiral limit; 
$\Delta_l(T) \equiv \lim_{m_l\rightarrow 0} \Delta_{l,s}(T)$ thus will 
become the standard order parameter for chiral symmetry restoration.
This normalized difference of condensates obtained in calculations with the
p4fat3 action on lattices with temporal extent $N_\tau =4,\; 6$ \cite{p4_eos}
and $8$ \cite{lat07} is shown in Fig.~\ref{fig:chipol}(left). 

In the
right hand part of this figure we show results for the renormalized
Polyakov loop obtained in the same set of calculations. As can be seen
the most rapid change in both quantities occurs in the same temperature
range also on lattices with temporal extent $N_\tau=8$ \cite{lat07}. 
A cut-off dependence, which shifts the transition region to smaller
temperatures, is visible in both observables. It, however, seems to be
small and correlated in both observables.

The rapid change in the chiral condensate reflected in Fig.~\ref{fig:chipol}
by the drop in $\Delta_{l,s}(T)$, of course, is correlated to a peak in the 
light quark chiral susceptibility, $\chi_{m,l}$, introduced in 
Eq.~\ref{msus}. This susceptibility actually is composed of two contributions, 
usually referred to as the connected and disconnected part,
\begin{equation}
\chi_{\rm tot}\equiv \chi_{m\, ,\, l} = \chi_{\rm disc} +\chi_{\rm con} \;
\end{equation}
with
\begin{eqnarray}
\chi_{\rm disc} &=& {1 \over 4 N_{\sigma}^3 N_{\tau}} \left\{
\langle\bigl( {\rm Tr} D_l^{-1}\bigr)^2  \rangle -
\langle {\rm Tr} D_l^{-1}\rangle^2 \right\} ~~, \cr
\chi_{\rm con} &=& - {1 \over 2} \sum_x \langle \,D_l^{-1}(x,0)
D_l^{-1}(0,x) \,\rangle~~.
\end{eqnarray}
Here $D_l$ denotes the staggered fermion
matrix for the light quarks. In Fig.~$5$ we show results for
the disconnected part of the light quark chiral susceptibility and the 
combined total susceptibility obtained in calculations performed on lattices 
of size $32^3\, 8$ with the asqtad and p4fat3 actions. This is compared
to the subtracted, normalized chiral condensate, $\Delta_{l,s} (T)$, calculated
with both actions on the same size lattices. As can be seen the peak in 
$\chi_{\rm disc}/T^2$ as well as $\chi_{\rm tot}/T^2$ obtained from 
calculations within both discretization schemes is in good agreement and 
corresponds well to the region of most rapid change in $\Delta_{l,s} (T)$. 

\section{Conclusions}

Studies of the thermodynamics of (2+1)-flavor QCD with a physical
value of the strange quark mass and almost physical values of the 
light quark masses have been performed at vanishing chemical
potential with two versions of ${\cal O}(a^2)$ improved 
staggered fermions, the asqtad and p4fat3 actions. Already on lattices
with temporal extent $N_\tau =6$ they yield a consistent description
of bulk thermodynamics, e.g. of the temperature dependence of energy 
density and pressure. This also holds true for the structure of the 
transition region and is confirmed through calculations closer to the
continuum limit performed on lattices with temporal extent 
$N_\tau=8$. ~These~ calculations~ yield 

\begin{figure}
\begin{minipage}[t]{0.5\linewidth}
\begin{center}
\vspace*{-0.4cm}
\epsfig{file=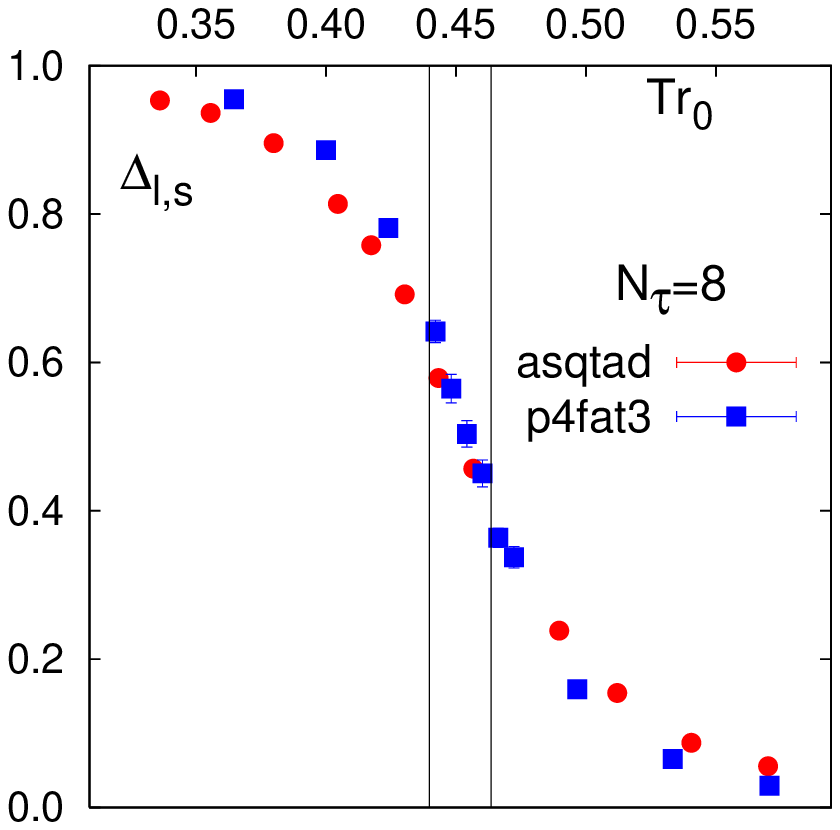,width=9.2cm}

\vspace*{-0.5cm}
\hspace*{-0.32cm}\epsfig{file=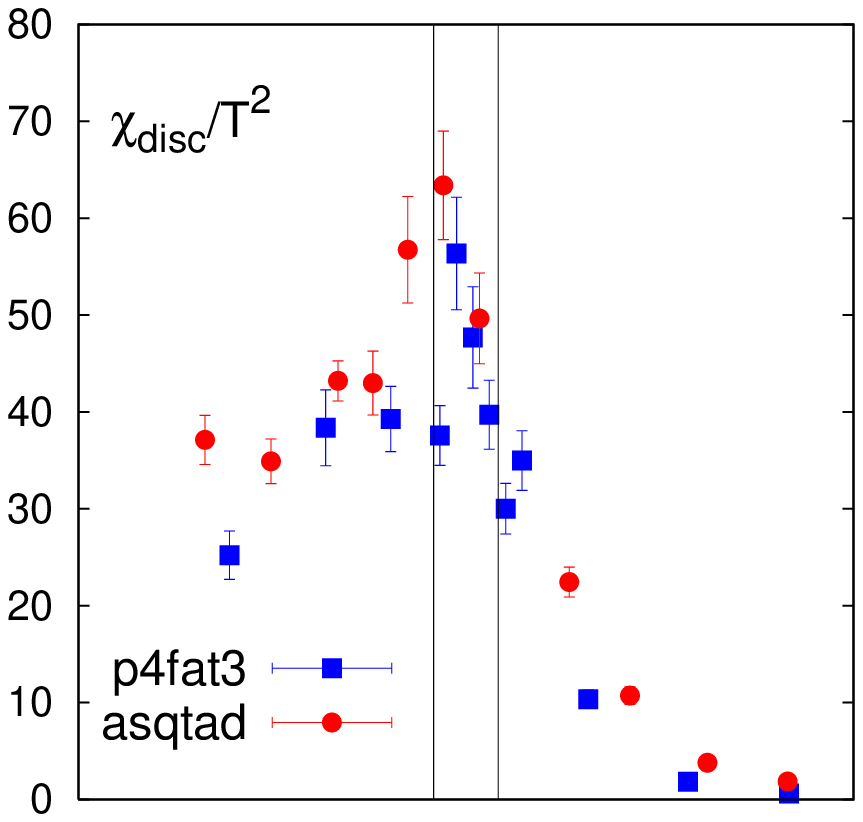,width=8.8cm}

\vspace*{-0.5cm}
\hspace*{-0.39cm}\epsfig{file=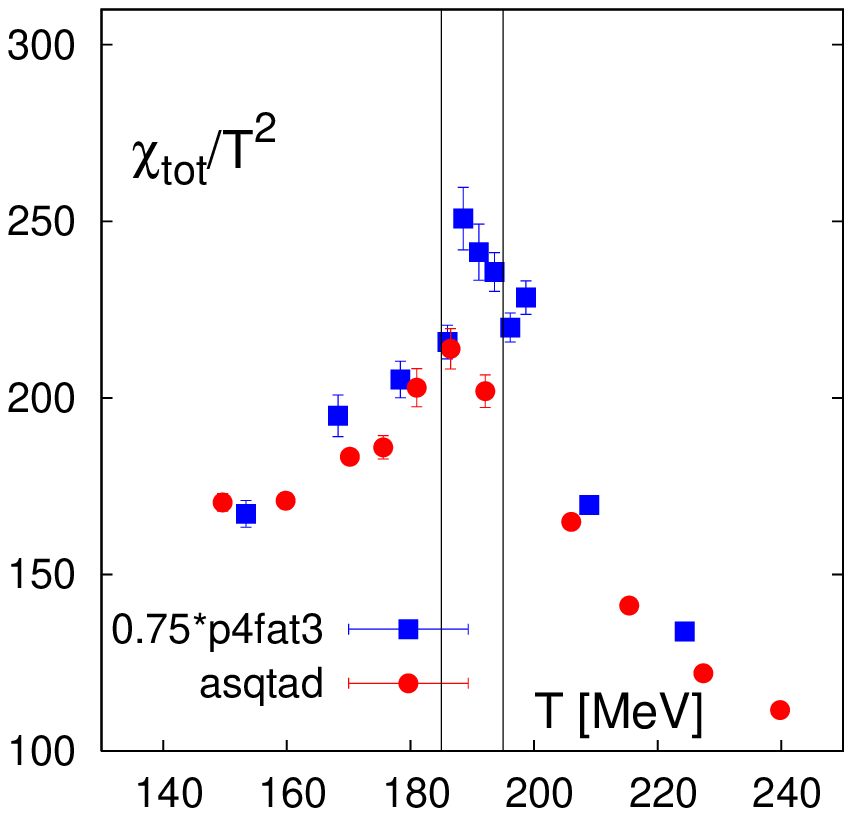,width=9.2cm}
\end{center}
\vspace*{-0.6cm}
\end{minipage}
\hspace{0.1cm} % To get a little bit of space between the figures
\begin{minipage}[t]{0.5\linewidth}
\caption{Subtracted finite temperature chiral condensates normalized
by the corresponding zero temperature quantity evaluated at the same
value of the cut-off (top), the disconnected part of the light quark chiral 
susceptibility (middle) and the total light quark chiral susceptibility
(bottom). All figures show preliminary results of the hotQCD Collaboration
obtained with two different ${\cal O}(a^2)$ improved staggered fermion
actions on lattices of size $32^3\, 8$ \cite{lat07}.  
}

\vspace*{0.4cm}
preliminary results for the transition temperature that 
may differ
by a few MeV, depending on the observable used to identify the transition
at non-zero quark mass values.
In particular, on the $N_\tau=8$ lattice no large differences in
the determination of the transition temperature arises from observables
related to deconfinement and chiral symmetry restoration respectively. 
The preliminary results of the hotQCD collaboration indicate that the 
crossover region for both deconfinement and chiral symmetry restoration 
lie in the range $T=(185$-$195$)~MeV for $N_\tau=8$ and $m_l/m_s =0.1$.

\end{minipage}
\label{fig:chi}
\end{figure}

\end{document}